\documentclass[12pt]{article}
\usepackage{amssymb,amsmath,cite,geometry}
\catcode`\@=11

\@addtoreset{equation}{section}
\def\theequation{\arabic{section}.\arabic{equation}}
     
\catcode`\@=11
\def\thesection{\arabic{section}}

\def\appendix{\setcounter{section}{0}
        \def\thesection{Appendix \Alph{section}}
        \def\theequation{\Alph{section}.\arabic{equation}}}
\def\section{\@startsection{section}{1}{\z@}{3.5ex plus 1ex minus
   .2ex}{2.3ex plus .2ex}{\large\bf}}
     
\long\def\@makefntext#1{\parindent 0cm\noindent
\hbox to 1em{\hss$^{\@thefnmark}$}#1}

\newcommand{\captionfonts}{\small}
\makeatletter  % Allow the use of @ in command names
\long\def\@makecaption#1#2{%
  \vskip\abovecaptionskip
  \sbox\@tempboxa{{\captionfonts #1: #2}}%
  \ifdim \wd\@tempboxa >\hsize
    {\captionfonts #1: #2\par}
  \else
    \hbox to\hsize{\hfil\box\@tempboxa\hfil}%
  \fi
  \vskip\belowcaptionskip}
\makeatother   % Cancel the effect of \makeatletter 

\begin{document}
\begin{titlepage}
\vspace{.5in}
\begin{flushright}
January 2011\\  %date
\end{flushright}
\vspace{.5in}
\begin{center}
{\Large\bf
 Extremal and nonextremal\\[1.3ex] Kerr/CFT correspondences}\\  %title
\vspace{.4in}
{S.~C{\sc arlip}\footnote{\it email: carlip@physics.ucdavis.edu}\\
       {\small\it Department of Physics}\\
       {\small\it University of California}\\
       {\small\it Davis, CA 95616}\\{\small\it USA}}
\end{center}

\vspace{.5in}
\begin{center}
{\large\bf Abstract}
\end{center}
\begin{center}
\begin{minipage}{5in}
{\small
I rederive the Kerr/CFT correspondence without first taking the near-horizon
extremal Kerr limit.  This method extends easily to nonextremal black holes,
for which the temperature and central charge behave poorly at the horizon
but the entropy remains finite.  A computation yields one-half of the standard
Bekenstein-Hawking entropy, with hints that the other half may be related 
to a conformal field theory at the inner horizon.  I then present an alternative 
approach, based on a stretched Killing horizon, in which the full entropy is 
obtained and the  temperature and central charge remain well-behaved even 
in the nonextremal case.
}
\end{minipage}
\end{center}
\end{titlepage}
\addtocounter{footnote}{-1}

The idea that black hole thermodynamics might be explained by a 
near-horizon conformal symmetry \cite{Strominger,Carlipa,Carlipb} has 
gained new currency with the recent discovery of an extremal Kerr/CFT 
duality \cite{Guica}.  A near-horizon conformal field theory is an 
attractive idea: it might help explain the ``universality'' of black hole 
entropy, and suggests a picture of the relevant degrees of freedom as 
Goldstone-like excitations arising from the conformal anomaly 
\cite{Carlipc}.  For the special case of an extremal Kerr black hole, 
the correspondence found in \cite{Guica} has additional attractive 
features, including a particularly simple central charge and perhaps a 
microscopic string theory realization \cite{Guicab}.  But although some 
progress has been made in extending this correspondence to near-extremal 
black holes \cite{Castro,Rasmussen}, the application to black holes far 
from extremality remains obscure.

The usual approach to the extremal Kerr/CFT correspondence begins by
approximating the extremal Kerr metric near the horizon as the ``near-horizon 
extremal Kerr,'' or NHEK, metric of Bardeen and Horowitz \cite{Bardeen}.
This form of the metric makes the near-horizon $\mathrm{AdS}_2$ structure
explicit, but does not generalize easily to the nonextremal case, where the 
near-horizon geometry is Rindler rather than anti-de Sitter.  But the NHEK 
limit, while convenient, should not be necessary: the horizon symmetry 
must surely be present in the full extremal Kerr metric, although perhaps 
better hidden.  

In the first section of this paper, I rederive the extremal Kerr/CFT 
correspondence without first going to the NHEK limit, using instead a 
stretched horizon formalism.  I then extend the calculation to an arbitrary 
stationary (3+1)-dimensional black hole, without requiring (near)-extremality. 
For  nonextremal black holes, the temperature and central charge computed 
in this manner behave badly at the horizon, but the combination that gives 
the entropy remains finite, yielding half the standard Bekenstein-Hawking 
entropy.  The factor of one-half can be traced to the presence of a single 
zero in the lapse function for a nonextremal black hole, in contrast to the 
double zero in the extremal case.  Since the extremal double zero comes 
from the merger of the inner and outer horizons, this suggests that the 
``missing'' entropy may be related to the inner horizon.

The bad behavior of the temperature and central charge at the horizon 
make this approach somewhat unsatisfying, however.   I therefore present 
an alternative near-horizon conformal field theory, living on a stretched 
Killing horizon.  This model, which is closely related to that of 
\cite{Carlipb}, has a finite temperature and central charge even as 
the stretched horizon approaches the true event horizon, while the 
singular behavior is shifted to a physical infinite blue shift at the horizon.
The Cardy formula, in both canonical and microcanonical form, now
yields the correct Bekenstein-Hawking entropy.

The covariant phase space approach to the central charge---first applied 
to black holes in \cite{Carlipb}, corrected and significantly extended in 
\cite{Koga,Silva}, and perfected in \cite{Barnich,Compere}---provides 
an elegant geometric description of the properties of a boundary 
conformal field theory.  Unfortunately, though, this formalism is a bit 
less transparent than one might hope for.  I  therefore use the more 
straightforward canonical ADM approach pioneered by Brown and 
Henneaux \cite{Brown}, which I briefly review in an appendix.

\section{Extremal Kerr/CFT without NHEK \label{s1}}

Let us begin with the extremal Kerr metric in Boyer-Lindquist coordinates,
written in ADM form as \cite{Frolov}
\begin{align}
ds^2 &= -N^2dt^2  + q_{ij}(dx^i + N^idt) (dx^j + N^jdt) \nonumber\\
       &= -N^2dt^2  + \frac{\Sigma}{\Delta}dr^2
      + \frac{A\sin^2\theta}{\Sigma}\left(d\varphi + N^\varphi dt\right)^2 
      + \Sigma\, d\theta^2 \ ,
\label{x1}
\end{align}
where $q_{ij}$ denotes the spatial metric on a constant time slice, and
\begin{align}
&\Delta = (r-r_+)^2, \qquad
\Sigma = r^2 + r_+{}^2\cos^2\theta, \qquad
A = (r^2 + r_+{}^2)^2 - (r-r_+)^2r_+{}^2\sin^2\theta ,
\nonumber\\
&N = \sqrt{\frac{\Sigma\Delta}{A}}  ,
 \qquad N^\varphi = \frac{2r_+{}^2r}{A} \ .
\label{x2}
\end{align}
The only nonvanishing component of the canonical momentum is
\begin{align}
\pi^{r\varphi} = \frac{\sqrt{q}}{2N}q^{rr}\partial_rN^\varphi \ .
\label{x3}
\end{align}
Near the horizon, the shift vector $N^\varphi$ can be expanded as
$N^\varphi \approx -\Omega_{\scriptscriptstyle\mathit{H}} 
+ \varepsilon$, where $\Omega_{\scriptscriptstyle\mathit{H}}$ is the
horizon angular velocity and the small parameter $\varepsilon$ is given by
\begin{align}
\varepsilon = (r-r_+)\partial_rN^\varphi\bigl|_{r=r_+} 
   = -\frac{(r-r_+)}{2r_+{}^2} \ .
\label{x4}
\end{align}

Under a diffeomorphism generated by a vector field $\xi^\mu$, the metric
transforms as
\begin{alignat}{3}
&\delta_\xi N 
        &&= {\bar\partial}_t\xi^\perp + {\hat\xi}^i\partial_i N \nonumber\\
&\delta_\xi N^i 
        &&= {\bar\partial}_t{\hat\xi}^i - N q^{ij}\partial_j\xi^\perp 
        + q^{ik}\partial_kN \xi^\perp + {\hat\xi}^j\partial_jN^i  \nonumber\\
&\delta_\xi q_{ij} 
        &&= q_{ik}\left(\partial_j{\hat\xi}^k 
                   -\frac{\partial_jN^k}{N}\xi^\perp\right)
        + q_{jk}\left(\partial_i{\hat\xi}^k 
                    -\frac{\partial_iN^k}{N}\xi^\perp\right)
                    +\frac{1}{N}\xi^\perp{\bar\partial}_t q_{ij} 
                    + {\hat\xi}^k\partial_kq_{ij}  \ , \label{a5}
\end{alignat}
where
\begin{align}
{\bar\partial}_t = \partial_t - N^i\partial_i  
     = \partial_t + \Omega_{\scriptscriptstyle\mathit{H}}\partial_\varphi 
     - \varepsilon\partial_\varphi
\label{a7}
\end{align}
is a convective derivative, and the quantities $(\xi^\perp,{\hat\xi}^i)$
are the ``surface deformation parameters'' (\ref{a6}) that appear in the 
ADM Hamiltonian \cite{Brown,Teitelboim}.  Note that for any function of 
the NHEK angular coordinate 
$\phi = \varphi-\Omega_{\scriptscriptstyle\mathit{H}}t$, 
\begin{align}
{\bar\partial}_t f = -\varepsilon\partial_\varphi f  \ .
\label{a8}
\end{align}

For extremal black holes, the horizon $\mathcal{H}$ is infinitely far 
from any stationary observer, in the sense that the proper distance 
from $r_+$ to any point $r>r_+$ is infinite.  The NHEK approach 
therefore treats boundary conditions at the horizon as asymptotic fall-off
conditions.  In Boyer-Lindquist coordinates, on the other hand, the
coordinate distance to the horizon is finite, and we must instead impose 
boundary conditions at $r=r_+$.  As usual, it is difficult to do this 
precisely at the horizon.  In the ADM coordinates (\ref{x1}), $N$ 
goes to zero at $\mathcal{H}$, and the metric becomes singular.  One 
could, of course, choose coordinates that are well-behaved at the horizon, 
but the real problem is more general: the horizon is a null surface, and 
the presence of second class constraints makes boundary conditions at 
such a surface extremely complicated.  I will therefore impose boundary 
conditions at a ``stretched horizon'' $\mathcal{H}_s$, and then take 
the limit as $\mathcal{H}_s$ approaches the true horizon $\mathcal{H}$.

As we shall later in this paper, there is more than one way to stretch 
a horizon.  It turns out the NHEK boundary conditions of \cite{Guica} 
correspond to fixing the angular velocity $N^\varphi$, or equivalently 
the parameter $\varepsilon$, thus determining a surface that rotates at a 
constant angular velocity 
$\Omega_s = \Omega_{\scriptscriptstyle\mathit{H}} - \varepsilon$ 
that differs slightly from $\Omega_{\scriptscriptstyle\mathit{H}}$.  
Indeed, if $\varepsilon$ is fixed, it is easy to see from (\ref{a5}) and (\ref{a8}) 
that $\delta_\xi N^r$ will be small as long as $\xi^r$ is a function of 
$\varphi-\Omega_{\scriptscriptstyle\mathit{H}}t$.  Then
\begin{align}
\delta_\xi N^\varphi = {\bar\partial}_t{\hat\xi}^\varphi 
     - N^2q^{\varphi\varphi}\partial_\varphi\xi^t 
     + {\hat\xi}^r\partial_rN^\varphi = 0
\label{a9}
\end{align}
has a solution
\begin{align}
{\hat\xi}^r = (r-r_+)\partial_\varphi{\hat\xi}^\varphi, \qquad
\xi^t = \mathcal{O}(r-r_+) \ .
\label{a10}
\end{align}
As in \cite{Guica}, these transformations allow $\mathcal{O}(1)$ 
changes in $q_{rr}$ and $q_{\varphi\varphi}$, but still lead to a
well-behaved variational principle; in particular, the variations of the 
conjugate variables $\pi^{rr}$ and $\pi^{\varphi\varphi}$ vanish 
as $r\rightarrow r_+$.  In fact, the group of diffeomorphisms (\ref{a10}) 
is equivalent to the asymptotic symmetry group of the NHEK metric found 
in \cite{Guica}:  not only is the algebra the same, but the vector fields 
themselves, when transformed to NHEK coordinates, match those of 
\cite{Guica} near the horizon.

We can now exploit the Cardy formula to determine the density of 
states.  This formula is most often seen in its microcanonical form 
\cite{Cardy,Cardyb}, and I discuss this form in \ref{aB}, but for our 
purposes the canonical version (see, for instance, section 8 of \cite{Bousso})
is more convenient.  For this, we need both the temperature $T$ and 
the central charge $c$.  The derivation of the temperature in \cite{Guica} 
did not involve the NHEK limit, and we can use that result directly:
\begin{align}
T = \frac{1}{2\pi} \ .
\label{x5}
\end{align}
I will return to this result in the next section, where a related but slightly 
different derivation is available for the nonextremal case.

To determine the central charge, we can turn to the expression (\ref{A8})  
of \ref{aA}, which gives the general central term of the surface 
deformation algebra in the canonical formalism.  For the group of 
deformations (\ref{a10}), it is easy to check that the only term that 
remains nonzero at the horizon is
\begin{align}
K[\xi,\eta] &= -\frac{1}{8\pi G}\int_{\partial\Sigma}d^2x
     \frac{\sqrt{\sigma}}{\sqrt{q}}\,n^k
     ({\hat\eta}_k\pi^{mn}D_m{\hat\xi}_n - {\hat\xi}_k\pi^{mn}D_m{\hat\eta}_n)
    \nonumber\\
    &= -\frac{1}{8\pi G}\int_{\mathcal{H}_s}d^2x\sqrt{\sigma}\, n^r
    q_{rr}{\hat\xi}^r \left(\frac{1}{2N}q^{rr}\partial_rN^\varphi\right)
    q_{rr}\partial_\varphi{\hat\eta}^r
    - ({\hat\xi}\leftrightarrow{\hat\eta})\nonumber\\
    &= -\frac{1}{16\pi G}\int_{\mathcal{H}_s}d^2x\sqrt{\sigma}\, 
    \frac{n_r}{N}\partial_rN^\varphi (r-r_+)^2
    \left(\partial_\varphi{\hat\xi}^\varphi \partial_\varphi^2{\hat\eta}^\varphi
    - \partial_\varphi{\hat\eta}^\varphi \partial_\varphi^2{\hat\xi}^\varphi\right)
    \ .  \label{x6}
\end{align}
But near the horizon,
\begin{align}
\frac{n_r}{N} = \frac{\sqrt{A}}{\Delta} \approx \frac{2r_+{}^2}{(r-r_+)^2}, 
\qquad \partial_rN^\varphi \approx -\frac{1}{2r_+{}^2} ,
\label{x7}
\end{align}
so
\begin{align}
K[\xi,\eta]  &= \frac{1}{16\pi G}\int_{\mathcal{H}_s}d^2x\sqrt{\sigma}\, 
\left(\partial_\varphi{\hat\xi}^\varphi \partial_\varphi^2{\hat\eta}^\varphi
    - \partial_\varphi{\hat\eta}^\varphi \partial_\varphi^2{\hat\xi}^\varphi\right)
   \nonumber\\
   &= \frac{1}{16\pi G}\frac{\mathcal{A}}{2\pi}\int d\varphi\,
\left(\partial_\varphi{\hat\xi}^\varphi \partial_\varphi^2{\hat\eta}^\varphi
    - \partial_\varphi{\hat\eta}^\varphi \partial_\varphi^2{\hat\xi}^\varphi\right)
    \ ,
\label{x8}
\end{align}
where $\mathcal{A} = \int d^2x\sqrt{\sigma}= 8\pi r_+{}^2$ is the horizon
area and I have used the fact that the metric near the horizon is independent 
of $\varphi$.

We can recognize (\ref{x8}) as the central term in a Virasoro algebra 
 \cite{Francesco} with central charge 
\begin{align}
c = 48\pi \frac{1}{16\pi G}\frac{\mathcal{A}}{2\pi}
   = \frac{3\mathcal{A}}{2\pi G} = 12J \ ,
\label{x10}
\end{align}
precisely matching the NHEK central charge of \cite{Guica}.  It is 
amusing to note that this quantity also matches the old near-horizon 
results of \cite{Carlipa,Carlipb}.  The canonical version of the Cardy 
formula then yields an entropy
\begin{align}
S = \frac{\pi^2}{3}cT = 2\pi J = \frac{\mathcal{A}}{4G} \ ,
\label{x11}
\end{align}
the standard Bekenstein-Hawking entropy.  

\section{Nonextremal black holes \label{s2}}

Extremality played a very small role in the preceding section, and it is 
straightforward to generalize the construction to the nonextremal Kerr 
black hole.  In fact, we can go farther, and consider an arbitrary 
stationary black hole.   Near the horizon, the metric of a stationary 
nonextremal (3+1)-dimensional black hole can always be written 
in the ADM form
\begin{align}
ds^2 = -N^2dt^2  + d\rho^2 
      + q_{\varphi\varphi}\left(d\varphi + N^\varphi dt\right)^2 
      + q_{zz}dz^2 \ ,
\label{a1}
\end{align}
where $\rho$ is the proper distance from the horizon and $z$ is,
for example, $\cos\theta$.   Independent of any field equations, the
requirement that the curvature be finite at the horizon sharply restricts 
the behavior of the metric: it must have an expansion of the form
\cite{Medved}
\begin{alignat}{3}
&N =  \kappa_{\scriptscriptstyle\mathit{H}}\rho 
       + \frac{1}{3!}\kappa_2(z)\rho^3 + \dots \qquad\quad
&& q_{\varphi\varphi} = [q_H]_{\varphi\varphi}(z) 
                    +  \frac{1}{2}[q_2]_{\varphi\varphi}(z)\rho^2 + \dots
       \nonumber\\[1ex]
&N^\varphi = -\Omega_{\scriptscriptstyle\mathit{H}} 
       - \frac{1}{2}\omega_2(z)\rho^2 + \dots 
&&q_{zz} = [q_H]_{zz}(z) + \frac{1}{2}[q_2]_{zz}(z)(\rho^2) + \dots \ ,
\label{a2}  
\end{alignat}
where the surface gravity $\kappa_{\scriptscriptstyle\mathit{H}}$ and 
the horizon angular velocity $\Omega_{\scriptscriptstyle\mathit{H}}$
are constants.  The only nonvanishing component of the canonical momentum
$\pi^{ij}$ at the horizon is
\begin{align}
\pi^{\rho\varphi}  
    =  -\frac{\omega_2}{2\kappa_{\scriptscriptstyle\mathit{H}}}\sqrt{q} 
   + \mathcal{O}(\rho^2) \ .
\label{a3}
\end{align}

As in the preceding section, we impose boundary conditions that 
$N^\varphi = -\Omega_{\scriptscriptstyle\mathit{H}} + \varepsilon$ 
is fixed at the stretched horizon $\mathcal{H}_s$.  Equation (\ref{a9}) 
now has a solution\footnote{The factor of $1/2$ difference between 
(\ref{ax1}) and (\ref{a10}) is a coordinate artifact, coming from the fact 
that $\rho\sim(r-r_+)^{1/2}$.}
\begin{align}
{\hat\xi}^\rho = -\frac{\rho}{2}\partial_\varphi{\hat\xi}^\varphi, \qquad 
\xi^t = \mathcal{O}(\rho) \ ,
\label{ax1}
\end{align}
with ${\hat\xi}^\varphi$ a function of the corotating coordinate 
$\varphi-\Omega_{\scriptscriptstyle\mathit{H}}t$.  As in the extremal 
case,  these diffeomorphisms preserve a sufficient set of boundary data 
on the stretched horizon.
 
Again as in the preceding section, we can use  (\ref{A8})  to determine the 
central charge.  The result is now
\begin{align}
K[\xi,\eta] &= -\frac{1}{8\pi G}\int_{\partial\Sigma}d^2x\frac{\sqrt{\sigma}} 
     {\sqrt{q}}\, n^k({\hat\eta}_k\pi^{mn}D_m{\hat\xi}_n 
           - {\hat\xi}_k\pi^{mn}D_m{\hat\eta}_n)
    \nonumber\\
    &= \frac{\varepsilon}{32\pi G\kappa_{\scriptscriptstyle\mathit{H}}}
     \frac{\mathcal{A}}{2\pi}\int d\varphi\, 
     (\partial_\varphi{\hat\xi}^\varphi \partial_\varphi^2{\hat\eta}^\varphi
     -\partial_\varphi{\hat\eta}^\varphi \partial_\varphi^2{\hat\xi}^\varphi)
     \ .  \label{a15} 
\end{align}
This is again the central term of a Virasoro algebra, with central charge
\begin{align}
c = \frac{3\mathcal{A}}{2\pi G}
    \frac{\varepsilon}{2\kappa_{\scriptscriptstyle\mathit{H}}} \ .
\label{ax2}
\end{align}

To determine the appropriate temperature to use in the Cardy formula, 
we can adapt the arguments of \cite{Guica}.  First, as usual, the Hawking 
temperature is 
\begin{align}
T_H = \frac{\kappa_{\scriptscriptstyle\mathit{H}}}{2\pi} \ .
\label{a11}
\end{align}
As noted in \cite{Guica}, however, this is not the relevant temperature for
our conformal algebra.  The Frolov-Thorne vacuum is annihilated by modes
with a coordinate dependence $e^{-i\omega t + im\varphi}$.  On our
stretched horizon, on the other hand, the relevant angular coordinate is
${\tilde\varphi} = \varphi - (\Omega_{\scriptscriptstyle\mathit{H}} 
+ \varepsilon)t$.  The Frolov-Thorne modes thus become
\begin{align}
\Phi_{m\omega} \sim e^{-i{\tilde\omega} t + im{\tilde\varphi}} \quad
\text{with}\  \ {\tilde\omega} 
    = \omega - m (\Omega_{\scriptscriptstyle\mathit{H}} + \varepsilon) ,
\label{a12}
\end{align}
with a Boltzmann factor
$
e^{-\beta(\omega - m\Omega_{\scriptscriptstyle\mathit{H}})} 
= e^{-\beta({\tilde\omega} - m\varepsilon)} 
$
The relevant temperature for $\tilde\varphi$ modes is thus
\begin{align}
T = \frac{T_H}{\varepsilon} 
    = \frac{\kappa_{\scriptscriptstyle\mathit{H}}}{2\pi\varepsilon} \ .
\label{a14}
\end{align}

The central charge (\ref{ax2}) and temperature (\ref{a14}) are poorly
behaved in the horizon limit $\varepsilon\rightarrow0$.  The entropy, 
however, is not.  Indeed, inserting the central charge and temperature 
into the canonical Cardy formula, we obtain 
\begin{align}
S = \frac{\pi^2}{3}cT = \frac{\mathcal{A}}{8G} \ ,
\label{a18}
\end{align}
one-half of the Bekenstein-Hawking entropy.

Mathematically, the missing factor of two can be traced back to the fact
that $N^2$ has a double zero at the horizon for an extremal black hole,
but only a single zero for a nonextremal black hole.  Indeed, for the 
nonextremal version of (\ref{x6}), the integrand contains a factor of
\begin{align}
\frac{n_r}{N}(r-r_+) = \frac{n_r}{2\partial_r N} 
    = \frac{1}{2\kappa_{\scriptscriptstyle\mathit{H}}} \ ,
\label{ax3}
\end{align}
where the factor of two comes from the fact that $N\sim\sqrt{r-r_+}$.
For the extremal black hole, $N\sim r-r_+$, and this factor is not present.

This observation suggests that the missing entropy could come from a 
conformal field theory at the inner horizon.  This is at least technically 
true, in the sense that the double zero of the extremal lapse function 
comes from the merger of single zeros at the inner and outer horizons.  
One might worry, though, both about the instability of the inner horizon 
\cite{Ori} and about the physical question of whether thermodynamic 
properties of a black hole can depend on a region that is causally 
disconnected from the exterior.

A possible alternative would be to find a second Virasoro algebra at the 
(outer) stretched horizon $\mathcal{H}_s$, describing a second set of 
modes.   While I cannot rule out the existence of such an algebra, I have
not succeeded in finding one.  In particular, the requirement of two
\emph{commuting} Virasoro algebras is a particularly strong one.  It may 
be, as Castro et al.\ have argued \cite{Castrob}, that one must search for 
a hidden nongeometric symmetry to obtain the remaining modes.

\section{More than one way to stretch a horizon \label{s3}}
 
The approach of the preceding sections has reproduced the NHEK 
expression for the entropy of an extremal black hole.   We have seen, 
however, that the nonextremal results are less convincing, both because 
of the missing factor of two in the entropy and because of the poor 
behavior of the temperature and the central charge at the horizon.    It 
is therefore worth looking for an alternative conformal description.

As noted above, there is more than one way to stretch a horizon.  The 
approach so far has been based on the fact that the horizon has a 
constant angular velocity $\Omega_{\scriptscriptstyle\mathit{H}}$; 
our stretched horizon has been a surface with a slightly different angular 
velocity $\Omega_s = \Omega_{\scriptscriptstyle\mathit{H}} - \varepsilon$.  
An alternative  approach, closer to that of \cite{Carlipa,Carlipb,Silva,Park}, 
is to note that the horizon of a stationary black hole\footnote{The 
analysis is local, so one need not require that the spacetime be globally 
stationary.  In fact, a general isolated horizon with a stationary 
neighborhood is a Killing horizon \cite{Date}.} is a Killing horizon:
that is, it admits a Killing vector 
$
\chi^a = T^a + \Omega_{\scriptscriptstyle\mathit{H}}\Phi^a
$ 
that is null at $\mathcal{H}$ and is normal to $\mathcal{H}$.  We 
cannot, of course, demand that the stretched horizon be a Killing 
horizon, but we can ``stretch'' the Killing vector, by requiring that a 
new Killing vector
\begin{align} 
{\bar\chi}^a =  T^a + {\bar\Omega}\Phi^a
\label{c1}
\end{align}
be null at the stretched horizon.  For the ADM metric (\ref{a1}), 
this means
\begin{align}
{\bar\chi}^2 = 0 
       =-N^2 + q_{\varphi\varphi}(N^\varphi + {\bar\Omega})^2
       = -N^2 + q_{\varphi\varphi}
          (\Omega_{\scriptscriptstyle\mathit{H}} - {\bar\Omega})^2
       + \mathcal{O}(\rho^3) \ .
\label{c2}
\end{align}
From the asymptotic behavior (\ref{a2}), it is evident that to lowest order 
we are fixing the proper distance $\rho$ at $\mathcal{H}_s$, a procedure
much closer to that of the usual ``membrane paradigm'' \cite{Thorne}. 
Note that the small parameter
\begin{align}
{\bar\varepsilon} = \Omega_{\scriptscriptstyle\mathit{H}} 
- {\bar\Omega} \  ,
\label{c2x}
\end{align}
which measures the ``stretching'' of the Killing vector ${\bar\chi}^a$,
is  of now order $\rho$.  This contrasts with the parameter $\varepsilon$ of 
section \ref{s1}, which was of order $\rho^2$.  It remains true, however, 
that as ${\bar\varepsilon}\rightarrow0$, the stretched horizon approaches
the true horizon.  Note also that although ${\bar\Omega}$ characterizes
the ``stretched Killing vector,'' it is not the angular velocity of the
stretched horizon.  Rather, by (\ref{a2}), the stretched horizon has
an angular velocity 
$-N^\varphi(\rho_s) = \Omega_{\scriptscriptstyle\mathit{H}}
+  \mathcal{O}({\bar\varepsilon}^2)$.

We now require that the form (\ref{a1}) of the metric remain fixed 
at the stretched horizon---that is, that the lapse function $N$, the 
shift vector $N^i$, and the (absent) cross-term $q_{\rho\varphi}$ 
be unchanged to lowest order in $\rho$.  Since we are treating the 
stretched horizon as a boundary, we can separately specify the surface 
deformation parameters $(\xi^\perp,{\hat\xi}^i)$ and their first normal 
derivatives $(\partial_\rho{\hat\xi}^i,\partial_\rho\xi^\perp)$ at
$\mathcal{H}_s$.  Let us consider parameters $\xi^t$  of the form 
$\xi^t =  \xi^t(\varphi-{\bar\Omega}t)$, where for the moment we
leave $\bar\Omega$ arbitrary; it will later be fixed by imposing
(\ref{c2}).  From (\ref{a5}), we then find that at the stretched 
horizon,
\begin{alignat}{3}
&\delta_\xi N = 0 \quad && \Rightarrow\ \
   {\hat\xi}^\rho = -{\bar\varepsilon}\rho\partial_\varphi\xi^t 
    = -\rho{\bar\partial}_t\xi^t\nonumber\\
& \delta_\xi N^\rho = 0 && \Rightarrow\ \
    \rho\partial_\rho\xi^t = -\frac{{\bar\varepsilon}^2}
  {\kappa_{\scriptscriptstyle\mathit{H}}^2}\partial_\varphi{}^2\xi^t
   = -\frac{1}{\kappa_{\scriptscriptstyle\mathit{H}}^2}{\bar\partial}_t{}^2\xi^t
   \nonumber\\
&\delta_\xi N^\varphi = 0 && \Rightarrow\ \
   {\hat\xi}^\varphi = \frac{\kappa_{\scriptscriptstyle\mathit{H}}^2\rho^2}
   {\bar\varepsilon}q^{\varphi\varphi}\xi^t =  {\bar\varepsilon}\nu^2\xi^t\nonumber\\
&\delta q_{\rho\varphi}=0 && \Rightarrow\ \
    \rho\partial_\rho{\hat\xi}^\varphi 
    = {\bar\varepsilon}\rho^2q^{\varphi\varphi}\partial_{\varphi}{}^2\xi^t 
    - \omega_2\rho^2\xi^t
    = \frac{{\bar\varepsilon}}{\kappa_{\scriptscriptstyle\mathit{H}}^2}
    {\bar\partial}_t{}^2\xi^t - \omega_2\rho^2\xi^t
\label{c3}
\end{alignat}
where
\begin{align}
   \nu^2 =  \frac{\kappa_{\scriptscriptstyle\mathit{H}}^2\rho^2}
  {q_{\varphi\varphi}{\bar\varepsilon}^2}
   = \frac{N^2}{q_{\varphi\varphi}(\Omega_{\scriptscriptstyle\mathit{H}}
         - {\bar\Omega})^2} \ .
\label{c4x}
\end{align}
We can also consistently require
\begin{align}
\delta_\xi g_{\rho\rho}=0 \quad \Rightarrow\ \ \partial_\rho{\hat\xi}^\rho=0
\label{c3x}
\end{align}
at $\mathcal{H}_s$, although this condition will not be needed for what follows.

Under the surface deformation brackets (\ref{A4}), it is now straightforward 
to check that to lowest order,
\begin{align}
\{\xi,\eta\}_{\scriptscriptstyle\mathit{SD}}^t 
    &= {\bar\varepsilon}( 1 + \nu^2) 
        (\xi^t\partial_\varphi\eta^t - \eta^t\partial_\varphi\xi^t) \nonumber\\
\{\xi,\eta\}_{\scriptscriptstyle\mathit{SD}}^\rho
     &= -{\bar\varepsilon}\rho\partial_\varphi
            \{\xi,\eta\}_{\scriptscriptstyle\mathit{SD}}^t \nonumber\\
\{\xi,\eta\}_{\scriptscriptstyle\mathit{SD}}^\varphi
     &= {\bar\varepsilon}\nu^2\{\xi,\eta\}_{\scriptscriptstyle\mathit{SD}}^t
     \ . \label{c4} 
\end{align}
These brackets form a Witt algebra---a Virasoro algebra with vanishing
central charge---but with a nontrivial normalization: 
\begin{align}
\xi^t_n = \frac{1}{(1+\nu^2){\bar\varepsilon}}\,
     e^{in(\varphi - {\bar\Omega}t)} \ .
\label{c5}
\end{align}
They  preserve the relations (\ref{c3}), providing a nontrivial consistency test.
Note that the wave vector $k_a$ determined by the exponent in (\ref{c5}) satisfies
\begin{align}
k^2 = n^2q^{\varphi\varphi}\left(1 - \frac{1}{\nu^2}\right) 
     + \mathcal{O}(\rho) \ ,
\label{c5x}
\end{align}
and is null when $\nu^2=1$, that is, when (\ref{c2}) is obeyed.

The nonvanishing contributions to the central term (\ref{A8}) are now
\begin{align}
K[\xi,\eta] &= -\frac{1}{8\pi G}\int_{\partial\Sigma}d^2x\sqrt{\sigma}\,
   n^k \Bigl[ (D_i{\hat\xi}_kD^i\eta^\perp  - D_i{\hat\xi}^iD_k\eta^\perp)
   - (\xi\leftrightarrow\eta)\Bigr] \nonumber\\
   &= -\frac{1}{8\pi G}\int_{\mathcal{H}_s}d^2x\sqrt{\sigma}\,  \Bigl[
   (q^{\varphi\varphi}\partial_\varphi{\hat\xi}^\rho \partial_\varphi\eta^\perp
   - \partial_\varphi{\hat\xi}^\varphi\partial_\rho\eta^\perp) 
   - (\xi\leftrightarrow\eta)\Bigr] \nonumber\\
   &= -\frac{\nu^2{\bar\varepsilon}^3}{4\pi G\kappa_{\scriptscriptstyle\mathit{H}}}
   \frac{\mathcal{A}}{2\pi}\int d\varphi\, 
   \left[ \partial_\varphi\xi^t\partial_\varphi{}^2\eta^t 
            - \partial_\varphi\eta^t\partial_\varphi{}^2\xi^t\right] \ .
\label{c6}
\end{align}
Taking into account the normalization (\ref{c5}), we can recognize this  
as the central term for a Virasoro algebra with 
central charge
\begin{align}
c = \frac{4\nu^2}{(1+\nu^2)^2}\frac{3\mathcal{A}}{2\pi G}
   \frac{{\bar\varepsilon}}{\kappa_{\scriptscriptstyle\mathit{H}}} \ .
\label{c7}
\end{align}

We can now fix the angular velocity $\bar\Omega$  by demanding that 
the modes be invariant under the translation along the ``stretched 
Killing vector'' $\bar\chi$, or, equivalently, that $\bar\chi$ be 
invariant under boundary diffeomorphisms, i.e., 
$\mathcal{L}_\xi{\bar\chi}=0$.   Then by (\ref{c2}), $\nu^2=1$, and
\begin{align}
c = \frac{3\mathcal{A}}{2\pi G}
       \frac{{\bar\varepsilon}}{\kappa_{\scriptscriptstyle\mathit{H}}} \ .
\label{c8}
\end{align}
It is interesting to note that this choice of $\nu^2$ gives the largest
possible  value of the central charge.  Note also that by (\ref{c5x}), this
is the unique choice for which $k^2=0$.  This is what one would expect 
from a two-dimensional conformal symmetry, since a null vector in 
Lorentzian signature translates to a holomorphic vector in Riemannian
signature.

As in section \ref{s2}, the temperature corresponding to the angular
dependence (\ref{c5}) is
\begin{align}
T = \frac{\kappa_{\scriptscriptstyle\mathit{H}}}{2\pi{\bar\varepsilon}}
\ . \label{c9}
\end{align}
The canonical Cardy formula then yields
\begin{align}
S = \frac{\pi^2}{3}cT = \frac{\mathcal{A}}{4G} \ ,
\label{c10}
\end{align}
as desired.  I show in \ref{aB} that the same result can be obtained from
the microcanonical version of the Cardy formula.

\section{Taming the central charge \label{s4}}

The approach of the preceding section has eliminated the ``factor of two''
problem.  As in section \ref{s2}, though, the temperature and the central 
charge for the nonextremal black hole are poorly behaved at the horizon.  
It may be, however, that this behavior really reflects something more 
physical, the infinite blue shift at the horizon relative to an observer 
outside the black hole.

To see this, let us replace the modes (\ref{c5}) with a slightly modified 
set,
\begin{align}
{\tilde\xi}^t_n 
    = \frac{1}{(1+\nu^2)}\,e^{in(\varphi - {\bar\Omega}t)/{\bar\varepsilon}}
\ , \label{d1}
\end{align}
chosen so that ${\bar\partial}_t{\tilde\xi}^t_n = in{\tilde\xi}^t_n$.  
The frequencies of these modes blow up at the horizon, but this is 
essentially the ordinary blue shift.  Indeed, a corotating observer (a 
ZAMO, or zero angular momentum observer) has a four-velocity 
$u^a = (T^a - N^\varphi\Phi^a)/N$, and by standard arguments will 
see a frequency $k_au^a$.  With the wave vector given by (\ref{d1}),
a simple computation yields $k_au^a = n/N(1+\mathcal{O}(\rho))$,
giving the standard blue shift near the horizon.

This choice of moding affects the algebra of deformations: because 
$\partial_\varphi$ is now $\mathcal{O}(1/{\bar\varepsilon})$, terms that 
were previously negligible are now important.  In particular, it is 
straightforward to check that
\begin{align}
\{{\tilde\xi},{\tilde\eta}\}_{\scriptscriptstyle\mathit{SD}}^t 
    =  ( 1 + \nu^2) ({\tilde\xi}^t{\bar\partial}_t{\tilde\eta}^t 
     - {\tilde\eta}^t{\bar\partial}_t{\tilde\xi}^t) 
    + \frac{1}{\kappa_{\scriptscriptstyle\mathit{H}}^2}
     ({\bar\partial}_t{\tilde\xi}^t{\bar\partial}_t{}^2{\tilde\eta}^t 
     - {\bar\partial}_t{\tilde\eta}^t{\bar\partial}_t{}^2{\tilde\xi}^t) \ ,
\label{d2}
\end{align}
which is no longer a Witt algebra.  As noted in \ref{aA}, though, the 
full surface deformation bracket must also include terms of the form 
$\{H[\xi],\eta\}$.  With the new moding, these are also no longer negligible.
The full brackets (\ref{A9}) become
\begin{align}
\{{\tilde\xi},{\tilde\eta}\}_{\scriptscriptstyle\mathit{full}}^t 
    &=  ( 1 + \nu^2) ({\tilde\xi}^t{\bar\partial}_t{\tilde\eta}^t 
     - {\tilde\eta}^t{\bar\partial}_t{\tilde\xi}^t) \nonumber\\
\{{\tilde\xi},{\tilde\eta}\}_{\scriptscriptstyle\mathit{full}}^\rho
     &= - \rho{\bar\partial}_t
            \{{\tilde\xi},{\tilde\eta}\}_{\scriptscriptstyle\mathit{full}}^t 
    \nonumber\\
\{{\tilde\xi},{\tilde\eta}\}_{\scriptscriptstyle\mathit{full}}^\varphi
     &= {\bar\varepsilon}\nu^2
     \{{\tilde\xi},{\tilde\eta}\}_{\scriptscriptstyle\mathit{full}}^t
     \ ,  \label{d4}
\end{align}
forming a standard Witt algebra and again preserving the relations (\ref{c3}). 

The central term (\ref{c6}) now becomes
\begin{align}
K[\xi,\eta] &= -\frac{1}{8\pi G}\int_{\partial\Sigma}d^2x\sqrt{\sigma}\, 
   n^k \Bigl[(D_i{\hat\xi}_kD^i\eta^\perp  - D_i{\hat\xi}^iD_k\eta^\perp)
   - (\xi\leftrightarrow\eta)\Bigr] \nonumber\\
   &= -\frac{\nu^2}{4\pi G\kappa_{\scriptscriptstyle\mathit{H}}}
   \frac{\mathcal{A}}{2\pi}\int d\varphi\, 
   \left[ {\bar\partial}_t\xi^t{\bar\partial}_t{}^2\eta^t 
            - {\bar\partial}_t\eta^t{\bar\partial}_t{}^2\xi^t\right] \ ,
\label{d5}
\end{align}
corresponding to a central charge\footnote{If the reciprocal of
$\bar\varepsilon$ is not an integer, the modes (\ref{d1}) are not periodic, 
and the boundary algebra will receive corrections.  These are 
of order $\bar\varepsilon$, though, and are negligible in the horizon limit.  
Alternatively, one may insist that the stretched horizon be chosen such 
that ${\bar\varepsilon} = 1/M$ for some large integer $M$.} of
\begin{align}
c = \frac{4\nu^2}{(1+\nu^2)^2}
   \frac{3\mathcal{A}}{2\pi G\kappa_{\scriptscriptstyle\mathit{H}}}
   =  \frac{3\mathcal{A}}{2\pi G\kappa_{\scriptscriptstyle\mathit{H}}} \ .
\label{d6}
\end{align}
This again matches the older results of \cite{Carlipa,Carlipb}.  The 
canonical Cardy formula, with the normal Hawking temperature, then 
yields
\begin{align}
S = \frac{\pi^2}{3}cT = \frac{\mathcal{A}}{4G} \ ,
\label{d7}
\end{align}
reproducing the standard Bekenstein-Hawking entropy.  Again, I show in
\ref{aB} that the same result can be obtained from the microcanonical
form of the Cardy formula.

It should be noted that although this approach gives a finite temperature
and central charge, the behavior is still somewhat singular at the horizon.
As observed above, the modes (\ref{d1}) are infinitely blue-shifted as
the stretched horizon approaches the true horizon.  In addition, the
$\rho$ derivatives in (\ref{c3}) blow up at the horizon.  This behavior
is also familiar, however: in terms of the usual ``tortoise coordinate''  
$r_* \sim \frac{1}{\kappa}\ln(\kappa\rho)$, we have $\rho\partial_\rho
\sim\partial_{r_*}$, so the apparent singular behavior corresponds
to the familiar smooth dependence of modes on $r_*$.  The condition
(\ref{c3x}) on $\xi^\rho$ is a bit more problematic, but as noted earlier,
this condition is not really needed.

\section{Conclusions}

The extremal black hole is special: its horizon is an infinite proper 
distance from any stationary observer, and has an asymptotically 
anti-de Sitter structure.  In retrospect, it is not so surprising that 
conventional asymptotic methods yield a conformal algebra that 
describes its states.

To perform a similar analysis in the nonextremal case, one traditionally
introduces a stretched horizon and imposes boundary conditions there.
We have seen that this procedure is not unique.  For the choice that most
closely resembles the NHEK approach to the extremal black hole, the
entropy appears to be split between the inner and outer horizons, which
converge only for the extremal black hole.  An alternative choice of 
``stretching the Killing horizon,'' on the other hand,  yields the full 
Bekenstein-Hawking entropy at the outer horizon.  It has been known for
some time that in such an approach, the relevant diffeomorphisms behave
poorly in the horizon limit \cite{Dreyer,Carlipd}.  But we have now seen 
that this is at least arguably a physical effect, resulting from the 
normal infinite blue shift at the horizon.

Ideally, one might hope to clarify these issues by looking at an appropriate
algebra of diffeomorphisms at a genuine horizon, without resorting 
to an intermediate stretched horizon.  Unfortunately, the constraint 
algebra becomes quite complicated on a null surface (see, for instance, 
\cite{Goldberg,Torre}), with an awkward set of second class constraints.  
Nevertheless, this avenue seems worth pursuing.

\vspace{1ex}
\begin{flushleft}
\large\bf Acknowledgments
\end{flushleft}

I would like to thank Jun-ichirou Koga for helpful comments.  Portions of 
this project were carried out at the Peyresq 15 Physics Conference with
the support of OLAM Association pour la Recherche Fundamentale, Bruxelles.
This work was supported in part by Department of Energy grant
DE-FG02-91ER40674.

\appendix
\section{Central terms in Hamiltonian gravity \label{aA}}

The symmetries of canonical general relativity are generated by the  
Hamiltonian
\begin{align}
H[\xi^\perp,{\hat\xi}^i] = \int d^3x \left(
         \xi^\perp\mathcal{H} + {\hat\xi}^i\mathcal{H}_i\right)
\label{A1}
\end{align}
with 
\begin{align}
\mathcal{H} = \frac{1}{\sqrt{q}}\left(\pi^{ij}\pi_{ij} - \pi^2\right)
     - \sqrt{q}\,{}^{(3)}\!R, \qquad
\mathcal{H}^i = -2D_j\pi^{ij} \ .
\label{A2}
\end{align}
Here, $q_{ij}$ is the spatial metric, $\pi^{ij}$ is the canonical momentum,
$D_j$ is the spatial covariant derivative compatible with $q_{ij}$, and 
$\mathcal{H}$ and $\mathcal{H}^i$ are the Hamiltonian and momentum
constraints; I use the sign conventions of \cite{Wald}, and units $16\pi G=1$.
On a manifold without boundary, these generators have Poisson brackets
\begin{alignat}{3}
&\left\{ H[\xi],q_{ij}\right\} &&=  -\frac{2}{\sqrt{q}}\xi^\perp\left(
     \pi_{ij} - \frac{1}{2}q_{ij}\pi\right)
     -  (D_i{\hat\xi}_j + D_j{\hat\xi}_i)\nonumber\\[.5ex]
&\left\{ H[\xi],\pi^{ij}\right\} &&= 
     \sqrt{q}\,\xi^\perp\left({}^{(3)}R^{ij} - \frac{1}{2}q^{ij}{}^{(3)}R\right) 
     +\frac{2}{\sqrt{q}}\,\xi^\perp \left(\pi^{ik}\pi_k{}^j 
                - \frac{1}{2}\pi\pi^{ij}\right) \nonumber\\
    &&& -\frac{1}{2}\frac{1}{\sqrt{q}}\,\xi^\perp q^{ij}\left(\pi_{mn}\pi^{mn} 
               - \frac{1}{2}\pi^2\right) 
    -\sqrt{q}\left(D^iD^j\xi^\perp - q^{ij}D_kD^k\xi^\perp\right)
    \nonumber\\[.5ex]
    &&&- D_k({\hat\xi}^k\pi^{ij}) + \pi^{ik}D_k{\hat\xi}^j + \pi^{jk}D_k{\hat\xi}^i
    \label{Ax} 
\end{alignat}
and
\begin{align}
\left\{ H[\xi], H[\eta]\right\} = H[\{\xi,\eta\}_{\scriptscriptstyle\mathit{SD}}]
\ , \label{A3}
\end{align}
where the surface deformation brackets \cite{Teitelboim} are
\begin{align}
&\{\xi,\eta\}_{\scriptscriptstyle\mathit{SD}}^\perp 
      = {\hat\xi}^iD_i\eta^\perp - {\hat\eta}^iD_i\xi^\perp \nonumber\\
&\{\xi,\eta\}_{\scriptscriptstyle\mathit{SD}}^i
      = {\hat\xi}^kD_k{\hat\eta}^i - {\hat\eta}^kD_k{\hat\xi}^i
      + q^{ik}\left(\xi^\perp D_k\eta^\perp - \eta^\perp D_k\xi^\perp\right) \ .
\label{A4}
\end{align}
The transformation (\ref{Ax}) is not a diffeomorphism, but is equivalent 
on shell to the diffeomorphism generated by a vector field $\xi^\mu$, with
\begin{align}
\xi^\perp = N\xi^t, \qquad {\hat\xi}^i = \xi^i + N^i\xi^t
\label{a6}
\end{align}

On a manifold with boundary, new complications arise.  The generators
(\ref{A1}), are not ``differentiable'' \cite{Regge}: a variation of the 
fields yields not only the standard functional derivative of the integrand,
but also a boundary term, typically singular, from partial integration.  
One must therefore add a boundary term to $H[\xi]$ to obtain a new 
generator
\begin{align}
{\bar H}[\xi] = H[\xi]  + B[\xi] \ ,
\label{A5}
\end{align}
where the new term $B[\xi]$ depends only on fields and parameters at the
boundary, and must be chosen to cancel the boundary terms in the variation
of $H[\xi]$. 

The specific form of $B[\xi]$ depends on  the detailed boundary conditions.
Even without knowing these, though, we can say a good deal about the 
Poisson brackets.  By definition, ${\bar H}[\xi]$ has a well-defined 
variation, with no boundary terms, so
\begin{align}
\left\{ {\bar H}[\xi],{\bar H}[\eta]\right\} = \int d^3x\left(
     \frac{\delta{\bar H}[\xi]}{\delta q_{ij}}
                            \frac{\delta{\bar H}[\eta]}{\delta \pi^{ij}}
   -\frac{\delta{\bar H}[\eta]}{\delta q_{ij}}
                            \frac{\delta{\bar H}[\xi]}{\delta \pi^{ij}}
   \right) \ .
\label{A6}
\end{align}
The functional derivatives in (\ref{A6}) can be read off from (\ref{Ax}).  
Inserting these and integrating by parts, and denoting the boundary 
normal and metric by $n^i$ and $\sigma_{ij}$, we obtain
\begin{align}
\left\{ {\bar H}[\xi],{\bar H}[\eta]\right\} 
    = {\bar H}[\{\xi,\eta\}_{\scriptscriptstyle\mathit{SD}}] 
    + K[\xi,\eta] \ ,
\label{A7}
\end{align}
where, restoring factors of $G$,
\begin{align}
K[\xi,\eta] = B[\{\xi,\eta\}_{\scriptscriptstyle\mathit{SD}}] 
    &-\frac{1}{8\pi G}\int_{\partial\Sigma}d^2x\sqrt{\sigma}\,n^k \Bigl[
    \frac{1}{\sqrt{q}}\pi_{ik}\{\xi,\eta\}_{\scriptscriptstyle\mathit{SD}}^i
    -\frac{1}{2}\frac{1}{\sqrt{q}}
               ({\hat\xi}_k\eta^\perp - {\hat\eta}_k\xi^\perp)\mathcal{H}
    \nonumber\\
    & + (D_i{\hat\xi}_kD^i\eta^\perp - D_i{\hat\eta}_kD^i\xi^\perp)
    - (D_i{\hat\xi}^iD_k\eta^\perp - D_i{\hat\eta}^iD_k\xi^\perp) \nonumber\\
    & + \frac{1}{\sqrt{q}}
     \left({\hat\eta}_k\pi^{mn}D_m{\hat\xi}_n
              - {\hat\xi}_k\pi^{mn}D_m{\hat\eta}_n\right)
    + (\xi^\perp\eta^i - \eta^\perp\xi^i){}^{\scriptscriptstyle(3)}\!R_{ik}
    \Bigr] \ . \label{A8} 
\end{align}
This is essentially the same expression as eqn.\ (4.43) of \cite{Brownb}.

To evaluate the central term (\ref{A8}), we still need  the boundary 
contribution $B[\{\xi,\eta\}_{\scriptscriptstyle\mathit{SD}}]$,
which will depend on our particular choice of boundary conditions.  For 
some purposes, though, the details are unnecessary.  In particular, 
suppose we find a Virasoro subalgebra of the group of surface deformations, 
with $\{\xi,\eta\}_{\scriptscriptstyle\mathit{SD}}=\xi\eta' - \eta\xi'$.  
The boundary term $B[\{\xi,\eta\}_{\scriptscriptstyle\mathit{SD}}] $
in (\ref{A8}) will then depend only on this combination.  A central 
term in a Virasoro algebra, on the other hand, looks like 
$\int\!d\varphi (\xi'\eta'' - \eta'\xi'')$, a form that cannot be built 
from $\xi\eta' - \eta\xi'$ and its derivatives.   If such a term occurs in 
$K[\xi,\eta]$, it must thus be a genuine central term.

One final subtlety can sometimes occur in the evaluation of the Poisson 
brackets (\ref{A3}).  So far, I have implicitly assumed that the surface 
deformation parameters $(\xi^\perp,{\hat\xi}^i)$ are independent of 
the canonical variables $(q_{ij},\pi^{ij})$.  If this is not the case, the 
Hamiltonian may have nontrivial Poisson brackets with the parameters
themselves.  In that event, the surface deformation brackets (\ref{A4}) 
become instead
\begin{align}
&\{\xi,\eta\}_{\scriptscriptstyle\mathit{full}}^\perp 
      = {\hat\xi}^iD_i\eta^\perp - {\hat\eta}^iD_i\xi^\perp 
      + \{H[\xi],\eta^\perp\} - \{H[\eta],\xi^\perp\} \nonumber\\
&\{\xi,\eta\}_{\scriptscriptstyle\mathit{full}}^i
      = {\hat\xi}^kD_k{\hat\eta}^i - {\hat\eta}^kD_k{\hat\xi}^i
      + q^{ik}\left(\xi^\perp D_k\eta^\perp - \eta^\perp D_k\xi^\perp\right)
      + \{H[\xi],{\hat\eta}^i\} - \{H[\eta],{\hat\xi}^i\} \ .
\label{A9x}
\end{align}

In particular, in sections \ref{s2}--\ref{s4} of this paper, the surface
deformation parameters depend on a coordinate $\rho$ which is  the proper 
distance to the horizon on a constant time slice.  This proper distance is 
metric-dependent: in terms of an arbitrary radial coordinate $r$, 
\begin{align}
\rho(x) = \int_{\mathcal{H}}^x \sqrt{g_{rr}}\,dr \ ,
\label{A10}
\end{align}
and hence, from (\ref{Ax}),
\begin{align}
\{H[\xi], F(\rho)\} = -{\hat\xi}^\rho\partial_\rho F + \mathcal{O}(\rho^2)
\ . \label{A11}
\end{align}
(I have omitted a term proportional to the extrinsic curvature $K_{rr}$, 
which vanishes at the horizon for the geometries considered here.)

For such metrics, the full surface deformation brackets (\ref{A9x}) are 
thus
\begin{align}
&\{\xi,\eta\}_{\scriptscriptstyle\mathit{full}}^\perp 
      = ({\hat\xi}^\rho \partial_\rho N)\eta^t 
                   - ({\hat\eta}^\rho \partial_\rho N)\xi^t
      + {\hat\xi}^\alpha \partial_\alpha\eta^\perp 
                   - {\hat\eta}^\alpha \partial_\alpha\xi^\perp
      \label{A9}\\
&\{\xi,\eta\}_{\scriptscriptstyle\mathit{full}}^i
      = {\hat\xi}^\alpha D_\alpha{\hat\eta}^i 
                    - {\hat\eta}^\alpha D_\alpha{\hat\xi}^i
      + ({\hat\xi}^\rho\partial_\rho N^i)\eta^t 
                    - ({\hat\eta}^\rho\partial_\rho N^i)\xi^t 
      + q^{ik}\left(\xi^\perp D_k\eta^\perp - \eta^\perp D_k\xi^\perp\right)
\ , \nonumber
\end{align}
where $\alpha$ ranges over the ``angular'' indices but not $\rho$.  The
disappearance of terms of the form ${\hat\xi}^\rho\partial_\rho{\hat\eta}^i$ 
has a simple geometrical explanation: it is simply a consequence of the fact
that the proper distance $\rho$ is a diffeomorphism-invariant quantity.

\section{Boundary terms near a horizon \label{aB}}

As noted in the preceding appendix, the form of the boundary term $B[\xi]$ 
in the Hamiltonian depends on the particular choice of boundary conditions.
While it is not essential to the main line of this paper, it is interesting to
work these out at the stretched horizon of a black hole with a metric of
the form (\ref{a1})--(\ref{a2}).  

The boundary terms in the variation of the Hamiltonian (\ref{A1}) are
\cite{Wald,Brownb}
\begin{align}
\delta H[\xi] = \dots - \int_{\partial\Sigma}d^2x\Bigl\{\sqrt{\sigma}\Bigl[ 
    \xi^\perp(&n^k\sigma^{\ell m} - n^m\sigma^{\ell k})D_m\delta q_{k\ell}
    \label{B1}\\
    &- D_m\xi^\perp(n^k\sigma^{\ell m} - n^m\sigma^{\ell k})\delta q_{k\ell}
     \Bigr] 
     + 2{\hat\xi}^i\delta\pi^\rho{}_i - {\hat\xi}^\rho\pi^{ij}\delta q_{ij} \Bigr\}
    \ . \nonumber
\end{align}
Ordinarily, the boundary metric $\sigma_{ij}$ is taken to be fixed, and 
the main contribution to the variation comes from the first term in (\ref{B1}).  
This leads to a boundary term $B[\xi]$ proportional to the extrinsic 
curvature $k$ of the boundary.  

For the black holes investigated here, on the other hand, the extrinsic 
curvature of the boundary---essentially the expansion---is proportional 
to $\rho$, and goes to zero at the horizon.  On the other hand, we do not 
require $\sigma_{ij}$ to be fixed at the horizon, so the second term in  
(\ref{B1}) gives a nonvanishing contribution.  In fact, it is fairly easy to 
see that for metrics of the form (\ref{a1})--(\ref{a2}),
\begin{align}
\delta H[\xi] = \dots - \int_{\partial\Sigma}d^2x\left\{\sqrt{\sigma}\,
     n^m\partial_m\xi^\perp\sigma^{\ell k}\delta\sigma_{k\ell} 
    - \xi^\perp n^m\partial_m (\sqrt{\sigma}\sigma^{\ell k}\delta\sigma_{k\ell})
    + 2{\hat\xi}^i\delta\pi^\rho{}_i\right\} + \mathcal{O}(\rho^2) \ .
\label{B2}
\end{align}
For variations that fix the normal $n^a$, the required boundary term is
thus
\begin{align}
B[\xi] = \frac{1}{8\pi G}\int_{\mathcal{H}_s}d^2x\left\{
   \sqrt{\sigma}\, n^m\partial_m\xi^\perp 
   - \xi^\perp n^m\partial_m\sqrt{\sigma}
   + {\hat\xi}^i\pi^\rho{}_i\right\} \ ,
\label{B3}
\end{align}
where I have restored factors of $G$.  In particular, diffeomorphisms that
satisfy condition (\ref{c3x})---that is, $\partial_\rho{\hat\xi}^\rho=0$ 
at the stretched horizon---leave $n^a$ fixed, and are consistent
with this choice of $B[\xi]$.  It may further be checked that for 
deformations of the form (\ref{c3}), the three-derivative terms in 
$\delta_\eta B[\xi]$ match $K[\xi,\eta]$ of eqn.\ (\ref{c6}), as they 
should \cite{Brown,Park}.

As noted in section \ref{s3}, though, condition (\ref{c3x}) is not
needed to obtain the central charge.  One might worry that without
this requirement, our choice of asymptotic symmetries could be inconsistent 
with the boundary term (\ref{B3}).  Here, however, we are rescued by a 
subtlety similar to the one discussed at the end of \ref{aA}: we must take 
into account the metric dependence of the proper distance $\rho$ and 
the coordinate position of the stretched horizon.  Explicitly, for a 
diffeomorphism  (\ref{c3}) generated by a vector field $\eta^\mu$, one 
has
\begin{align}
\delta_\eta \left[\sqrt{\sigma}\, n^m\partial_m\xi^\perp\right] &=
    (\delta_\eta\sqrt{\sigma})\, n^m\partial_m\xi^\perp 
    - (\partial_\rho\eta^\rho) \sqrt{\sigma}\, n^m\partial_m\xi^\perp
    \nonumber\\
    &+ (\eta^\rho\partial_\rho\sqrt{\sigma})\, n^m\partial_m\xi^\perp
    + \sqrt{\sigma}\, (\eta^\rho\partial_\rho n^m)\partial_m\xi^\perp
    + \sqrt{\sigma}\, n^m\partial_m(\eta^\rho\partial_\rho\xi^\perp)
    \nonumber\\
    &= (\delta_\eta\sqrt{\sigma})\, n^m\partial_m\xi^\perp
    + \eta^\rho\partial_\rho(\sqrt{\sigma}\, n^m\partial_m\xi^\perp) \ ,
\label{Bx}
\end{align}
where the ``extra'' $\eta^\rho\partial_\rho$ terms come from the
dependence (\ref{A10}) of $\rho$ on the 
metric.  The effect of the last term in (\ref{Bx}) is simply to move 
the argument of the integrand of (\ref{B3}) from $\rho$ to 
$\rho+ \eta^\rho$.  But the location of the stretched horizon at 
$\rho=\rho_s$ is also moved to $\rho + \eta^\rho=\rho_s$, so 
this produces no change in $B[\xi]$; only the $\delta_\eta\sqrt{\sigma}$
piece remains.

Given the boundary term (\ref{B3}), we can now use the microcanonical
version of the Cardy formula to check the results of this paper.  For the
modes (\ref{c5}) of section \ref{s3},
\begin{align}
B[\xi_0] = \frac{1}{8\pi G}\int_{\mathcal{H}_s}d^2x\,\sqrt{\sigma}
   \frac{\kappa_{\mathit{\scriptscriptstyle H}}}{(1+\nu^2){\bar\varepsilon}}
   = \frac{\kappa_{\mathit{\scriptscriptstyle H}}\mathcal{A}}
      {16\pi G{\bar\varepsilon}} \ .
\label{B4}
\end{align}
With the central charge (\ref{c8}), the microcanonical Cardy
formula then yields
\begin{align}
S = 2\pi\sqrt{\frac{cL_0}{6}} = \frac{\mathcal{A}}{4G} \ .
\label{B5}
\end{align}
Similarly, for the modes (\ref{d1}) of section \ref{s4}, 
\begin{align}
B[{\tilde\xi}_0] = \frac{1}{8\pi G}\int_{\mathcal{H}_s}d^2x\,\sqrt{\sigma}
   \frac{\kappa_{\mathit{\scriptscriptstyle H}}}{(1+\nu^2)}
   = \frac{\kappa_{\mathit{\scriptscriptstyle H}}\mathcal{A}} {16\pi G} \ ,
\label{B6}
\end{align}
and the microcanonical Cardy formula with the central charge (\ref{d6}) 
yields
\begin{align}
S = 2\pi\sqrt{\frac{cL_0}{6}} = \frac{\mathcal{A}}{4G} \ .
\label{B7}
\end{align}

I do not know how to do a corresponding computation for the NHEK-type
symmetry of section \ref{s2}.  It is interesting to note, though, that for
the extremal case, a choice $L_0 = c/24$ gives the correct entropy 
(\ref{x11}).

\end{document}